# The Influence of Macroscopic Pedestrian Structures on Train Boarding Efficiency

Rabia I. Kodapanakkal[a], Caspar A.S. Pouw[b,c], Antal Haans[a], Jaap Ham[a], Gunter Bombaerts[d], Alessandro Corbetta[b], Andrej Dameski[d], Andreas Spahn[d], Federico Toschi[b,e]

[a]Human-Technology Interaction Group, Eindhoven University of Technology, Eindhoven 5600 MB, The Netherlands

[b]Department of Applied Physics, Eindhoven University of Technology, Eindhoven 5600 MB, The Netherlands

[c]ProRail BV, Moreelsepark 2, Utrecht 3511 EP, The Netherlands

[d]Philosophy and Ethics Group, Eindhoven University of Technology, Eindhoven 5600 MB, The Netherlands

[e]CNR-IAC, Rome, Italy

**Declaration of Interests:** No competing interests exist.

**Funding:** This work is part of the HTSM-AES research programme "HTCrowd: a high-tech platform for human crowd flows, monitoring, modeling and nudging" with project number 17962, and the MVI top up grant "SRCrowd: Individual and collective agency in Socially Responsible nudging of Crowds" with project number 18754, both financed by the Dutch Research Council (NWO).

Correspondence can be addressed to Antal Haans at a.haans@tue.nl

Draft date: 2023 Sep 07

**Abstract**

A deeper understanding of pedestrian dynamics is essential to improve crowd flows in public spaces such as train stations. It is essential to understand both the physical and the psychological processes present in this context. However, current research on train boarding behavior is limited in scope and mainly focuses on how group level variables such as number of boarders/deboarders influence train boarding efficiency. Viewing pedestrian dynamics through a psychological lens is important for a detailed understanding of the train boarding context and to recognize target areas for improving crowd flows. At Dutch train stations, boarders follow a social norm of waiting at the train door until deboarding is complete. Although people generally adhere to this norm, the way it is executed may not be optimal for deboarding efficiency. We investigate how waiting boarders form a deboarding channel (a corridor where deboarders exit the train) which is a macroscopic structure formed by pedestrians, and how this channel in turn influences the efficiency of deboarding. Analyzing a dataset with 3278 boarding events at Utrecht Centraal Station in the Netherlands from 2017 – 2020 (a subset of a trajectory dataset that captures 100,000 trajectories per day), we found that higher numbers of boarders and a higher ratio of boarders to deboarders, reduced the width of the deboarding channel, and a lower width was associated with lower deboarding efficiency. These results shift the focus from group level variables to identifying macroscopic structures that are formed when pedestrians interact within a social system and provide specific target areas where nudges/behavioral interventions could be implemented.

*Keywords: train boarding, deboarding efficiency, social norms, deboarding channel*

**The Influence of Macroscopic Pedestrian Structures on Train Boarding Efficiency**

Managing crowd flow is important to ensure a smooth flow of pedestrians in public spaces, such as train and metro stations. Often physical constraints such as bottlenecks hinder the smooth flow of pedestrians as it limits how many individuals can pass through (Seyfried et al., 2009). In addition to physical constraints, crowd flow at bottlenecks can also be affected by psychological factors such as stress, social norms, and motivation (Adrian et al., 2020). Therefore, to improve crowd flow at bottlenecks, it is important to understand pedestrian behavior both from a physical (Corbetta & Toschi, 2023) and psychological (Sieben et al., 2017) perspective.

At train stations, bottlenecks are commonly observed near the stairways/escalators and at the train doors where commuters enter and exit the train (Qu et al., 2019; Zhuang et al., 2017). Bottlenecks at the train door could result in inefficient boarding and deboarding and increase dwell times (total time that a train is halted at a railway platform) of the trains. Dwell times are also influenced by passenger behavior while boarding and deboarding trains (Kuipers et al., 2021). This increase in dwell time can result in accumulated delays across the trains' journeys and affect other trains' efficiency in the rail network (Christoforou et al., 2020). Since the number of passengers that use public transport is steadily growing and public transport is seen as a greener alternative to driving (Buehler, 2009), the efficiency of railway networks becomes very important. It is thus essential to understand the factors that influence the efficiency of the train boarding process.

In this paper, we take both physical and psychological perspectives to investigate bottlenecks in the train boarding process at Dutch train stations. We combine pedestrian trajectory data in real-life settings with psychological perspectives, focusing on social norms, to better understand pedestrian behavior. Specifically, we investigate how a deboarding channel (macroscopic structure) formed by groups of boarders affects the efficiency of deboarders exiting the train. This is an important aspect to study because it tells us more about the psychological processes through which passenger number influences efficiency. It also provides a more in-depth view of how two groups – boarders and deboarders – interact within a social system and how macroscopic structures like the beboarding channel emerge as a result of social norms governing this social system. Knowing more about this process puts researchers and practitioners in a better position to develop behavioral change interventions and nudges to improve train boarding efficiency as it allows them to empirically determine which macroscopic structures to target for improving efficiency.

**Background**

Research has by and large focused on two factors that can influence boarding efficiency and punctuality. The first factor relates to the characteristics of the physical environment such as the platform design, train door design, or train length (e.g., Coxon et al., 2009; Thoreau et al., 2016). The second factor relates to aspects associated with passengers' influence on the boarding process, such as how the number, ratio, and distribution of boarders and deboarders influence boarding efficiency (e.g., de Ana Rodriguez et al., 2016; Olsson & Haugland, 2004; Seriani et al., 2016). Although this latter research focuses on passenger influence, the process of how passengers actually affect efficiency is still unclear. We identify three gaps in the literature that our research seeks to address.

First, research has typically focused on identifying group level variables, such as number of boarders and ratio of boarders to deboarders, that are associated with efficiency and punctuality (Seriani et al., 2016). However, this association alone does not provide a good explanation of the process through which these group level variables affect efficiency. We argue that investigating macroscopic structural changes that result from changes in the number of boarders and/or deboarders can provide insights into the process of boarding and deboarding.

Second, prior research often does not separately measure the boarding and deboarding efficiency, so we do not know whether one process is more or less efficient than the other. Measures are often related to dwell time in total or total boarding time (Seriani & Fernandez, 2015). Some research does quantify boarding and deboarding rates separately (Harris et al., 2014) but does not investigate interactions between these two groups (boarding and deboarding groups) within the social system. For example, how the behavioral patterns and structures of waiting groups of boarders might specifically affect the efficiency of deboarders (and vice versa) has not been assessed. We argue that to understand the underlying structural changes that have an effect on the efficiency of boarding and deboarding, these processes should be considered separately since the influence these groups have on each other varies during both processes. For example, waiting boarders can exert influence on the deboarders by either providing or not providing enough space to them whereas deboarders generally do not influence the boarders as they leave the area around the train door before boarding commences.

Third, most earlier research was either conducted as laboratory experiments (Daamen et al., 2008; Seriani et al., 2019), simulations (Zhang et al., 2008) or manual observations at train stations (Li et al., 2020), all of which could have limitations of ecological validity and/or accuracy. Some recent research focusing on Swiss train stations does use real-time pedestrian trajectory data (Dell'Asin & Hool, 2018; Hänseler et al., 2015) but does not test the relationships between structures formed by boarding groups and how this affects efficiency. We propose that to investigate these structural behavioral patterns and their effect on efficiency accurately, using real-life data with high statistical power would be an advantage.

In the present research, we address these gaps. We use real-life data collected via sensors at a Dutch train station (Utrecht Centraal) and specifically investigate the macroscopic structure formed during the deboarding process as a result of the interaction between boarders and deboarders. We limit our focus to a single macroscopic structure, namely the deboarding channel formed as a result of waiting boarders providing space for the deboarders. We assess how group level variables like the boarding and deboarding group size affect the deboarding channel's width, and how the channel width in turn affects deboarding efficiency.

*Channel Width and Social Norms*

Boarders often form two groups, one on either side of the train door, to allow deboarders to exit the train (Dell'Asin & Hool, 2018; also see Figure 1). This formation gives rise to a macroscopic structure within which deboarders can move. This structure is in the form of a channel with a certain width and length. Although past research has investigated the number of boarders and ratio of boarders to deboarders affecting the efficiency of the overall boarding process (e.g., de Ana Rodriguez et al., 2016; Seriani et al., 2016), the formation of this channel can shed more light on how these factors affect the efficiency of the process. For example, with an increase in the number of boarders – and thus possible increase in the competition for seats – it is likely that boarders might want to wait as close to the door as possible and reduce the channel width as they move closer to the door.

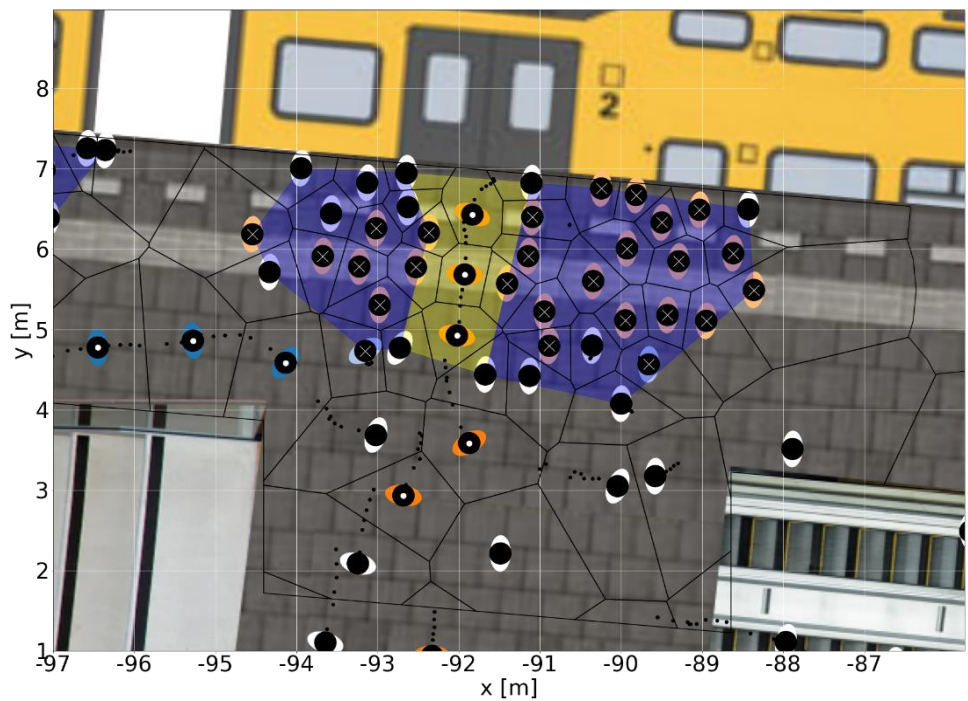

*Figure 1.* A visual representation of real-life data of train deboarding. Pedestrians are indicated with dots. Pedestrians in the blue areas are boarders waiting to board the train. Pedestrians in the yellow area (indicating the deboarding channel) are deboarders who are leaving the train. X- and y-axes indicate x- and y-positions of the pedestrians. Polygons around each pedestrian indicate the Voronoi area, or area available per person.

Since there are no explicit rules, and very limited physical constraints or directives (e.g., fences or signs) at Dutch train stations regarding what passengers should or should not do during boarding and deboarding, there appears to be some level of organization occurring during the boarding process. To understand what kind of macroscopic structure might emerge as a result of this organization in the context of Dutch train stations, we complement the physical data with a psychological framework. We propose that the deboarding channel formed during the boarding and deboarding process could be a result of people adhering to social norms. Social norms are unwritten rules of behavior that people conform to when they believe that others conform to them and others think it is the right thing to do (Bicchieri, 2016). One of the functions of social norms is that they are beneficial for cooperation and social order (Anderson & Dunning, 2014; Legros & Cislaghi, 2020). In the current research, we focus on the social norm that boarders wait around the train door for deboarders to leave the train. People feel like they are expected to follow this unwritten rule. This norm also contributes to cooperation and social order between boarders and deboarders and potentially improves the efficiency of the boarding and deboarding process.

However, it is not clear whether the way in which boarders execute this social norm is in fact efficient. Thus, an open question remains: *Although boarders follow a social norm and wait for deboarders to leave the train first, do they leave sufficient space for deboarders to ensure an efficient process?* To investigate this open question and to understand how the number of boarders and deboarders can influence the efficiency of deboarding via the deboarding channel, we ask the following questions. *1) Do the number of boarders and deboarders affect the dimensions of the deboarding channel formed? and 2) Does an increase or decrease in the dimensions of the channel respectively increase or decrease the efficiency* *of deboarding?* We expect that an increase in the number of boarders and an increase in the ratio of boarders to deboarders would decrease the width of the channel. We also expect that a decrease in the width of the channel would decrease the deboarding efficiency.

## Method

**Tracking Data Setup**

In the current work, we use pedestrian tracking data acquired from platform 5 of Utrecht Centraal station in The Netherlands. This is one of the most central and busiest stations in the Netherlands with approximately 188,933 passengers per day on average in the year 2022 ([Reizigersgedrag 2022, NS](#)). The pedestrian data is collected via commercial pedestrian tracking sensors which capture 3D stereo images at the rate of 10 frames per second. The data is processed to produce individual tracking data and this is done without compromising individual privacy. For more details on the tracking setup see Pouw et al. (2020).

**Inclusion Decisions**

For the purpose of this study, we focused only on a particular region of the platform and included doors that were within a distance of 1 meter from a particular sensor to minimize the number of false negative detections due to occlusions (cf. green area in Fig. 2). The platform width in the region was 3.5 meters. This also minimized the influence of obstacles on the platform and kept width of the platform constant to ensure that these factors wouldn't confound the explanation of our findings. Since we were only interested in the train doors that were used by passengers, we excluded doors that had no boarders or deboarders. We included data only from 2017 (when the sensors were installed) until 2020 (right before the Covid-19 pandemic started, diminishing the number of passengers greatly). The reason for this decision was to limit the dataset only to pre-pandemic boarding situations as we expected the patterns during the pandemic to be different and because we were mainly interested in patterns that were not restricted by social distancing rules which were applied during the pandemic. Since we were interested in the process of deboarding and what it is influenced by, the dataset only included the tracking data near the doors of the train and only the deboarding time was considered. Thus, the data included positions of passengers from the start of deboarding (when the first person leaves the specific train door) until the end of deboarding (when the last person leaves the specific train door). With these inclusion decisions, our final dataset consisted of 3278 deboarding events. The detailed process to localize door positions, and calculate parameters are reported in the following sub-sections.

**Algorithms**

To assess the boarding efficiency of a train, it is crucial to know the location of all its doors. However, the position of all the train doors is not known a priori and is affected by several factors, like the train length, rolling stock, and the location at which the train is stopped by its operator. Therefore, we need an accurate algorithm to infer the train door location based on the recorded pedestrian trajectories. The trajectory origins and destinations, measured during some time interval, disclose all the access points of the measurement domain. For instance, a group of trajectory destinations near the edge of the platform indicates the presence of a train door (red data points in Fig 2). Fitting the relative positions of the detected clusters with the physical distances between consecutive train doors ($\Delta x_1$ and $\Delta x_2$ from Fig. 2) yields an accurate localization of all the train doors, even allowing for the interpolation of train doors where only a few or no pedestrians (de)board the train.

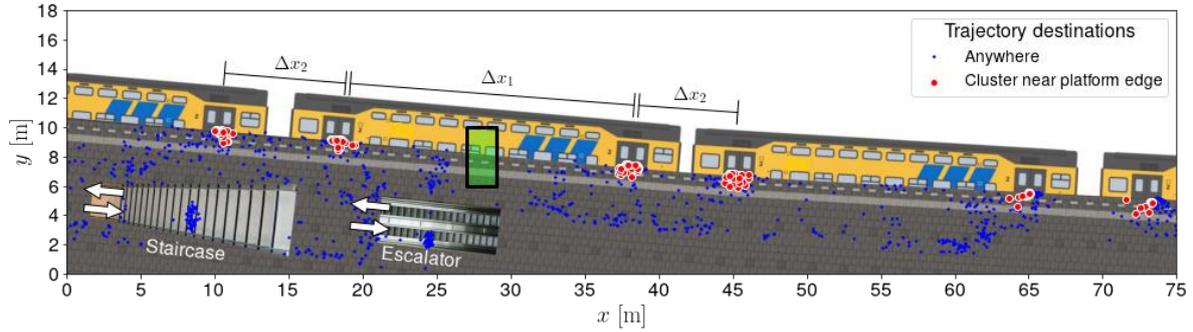

*Figure 2.* Overhead image of (a part of) the train platform at Utrecht CS (NL). We mark the location of the staircase and the escalators together with arrows to illustrate the direction of the crowd flows. We also impose the trajectory destinations that are clustered near the platform edge and which we use to infer the train door locations with a red color. Trajectory destinations in other locations are colored blue. We observe six clusters at regular intervals connected to the distances between contiguous train doors. This entails either the distance between doors of the same coach $\Delta x_1$, or doors of adjacent coaches $\Delta x_2$. Additionally, we highlight the range of train door locations that are considered in our analyses with a green square.

For the train boarding analysis, pedestrians were classified into one of the following three groups: (1) deboarding the train, (2) waiting to board the train, or (3) walking towards the train. Based on the origin and destination of a trajectory we classified if a pedestrian was deboarding (1), i.e. initial position close to a train door, or boarding (2 or 3), i.e. final position close to a train door. To further separate the boarding pedestrians into waiting (2) and walking (3) we used the instantaneous velocity of the trajectories. Pedestrians with a velocity higher than v > 0.4 m/s for a duration of at least t > 1 s (10 frames) were considered to be walking and therefore classified as walking toward the train (3).

During the deboarding process, we observed two separate clusters of pedestrians waiting to board the train (2) next to each train door, one cluster on each side of the train door, separated by a file of travelers deboarding the train (cf. Fig. 1). This separation enabled us to extract the left and right boarding clusters for each train door by employing a density-based clustering algorithm (DBSCAN) (Ester et al., 1995) that groups the boarding passengers that are closely packed together.

For every train door, the space available to deboarders was determined by the distance between the left and right boarding cluster. To find this distance, we computed, for both clusters, the convex hull that envelopes all the people part of the cluster and calculated the distance between the two convex hulls. For this calculation, the minimum distance between the two groups of waiting passengers i.e. the narrowest section of the channel was considered because this characterized the flow of the deboarders.

For every train door, we also calculated the total time taken for the deboarders to deboard the train. This duration was calculated as the time when the first deboarder was identified until the last deboarder left the train door. Thus, the following variables were extracted from the final dataset: number of boarders, number of deboarders, duration of deboarding, and the minimum width of the channel.

**Other Measures**

Using the number of boarders and deboarders, we calculated the proportion of boarders among all travelers (boarders and deboarders). This was calculated as:

$$\frac{no.\,of\,boarders}{(no.\,of\,boarders + no.\,of\,deboarders)}$$

Thus a proportion of 0 meant that for the particular door, there were no boarders and only deboarders. A proportion of 1 meant that there were only boarders and no deboarders. A proportion of 0.3 meant that among all travelers at a door, 30% of the travelers were boarders and 70% were deboarders. This measure gave an indication of how big or small the boarding groups were compared to deboarding groups. To estimate efficiency of the deboarding, we calculated deboarding flux as the total number of deboarders exiting the train per second and per meter of the door width:

$$\frac{no.\,of\,deboarders}{\frac{deboarding\,duration\,(s)}{train\,door\,width\,(m)}}$$

The width of all train doors in the dataset was 1.3 meters.

## Results

**Descriptives**

Across the 3278 deboarding events in the dataset, the number of waiting boarders ranged from 1 to 57 and the number of deboarders leaving the train door ranged from 1 to 62. The average deboarding flux across the events was 0.65 *persons/s/m* with a standard deviation of 0.17 *persons/s/m*. The average minimum channel width across the events was 1.37 *m* with a standard deviation of 0.36 *m*. Note that the actual average channel width is lower than this measured width taking into account the physical body size of people at the edge of the corridor. The two main independent variables – number of boarders and proportion of boarders – are correlated with each other with a Pearson correlation coefficient of r = 0.47. This means that on average when the number of boarders is higher, their proportion is also higher. The number of boarders and deboarders are correlated with a correlation coefficient of $r = 0.17$. See Figures 3 and 4 for visualizations of the measures.

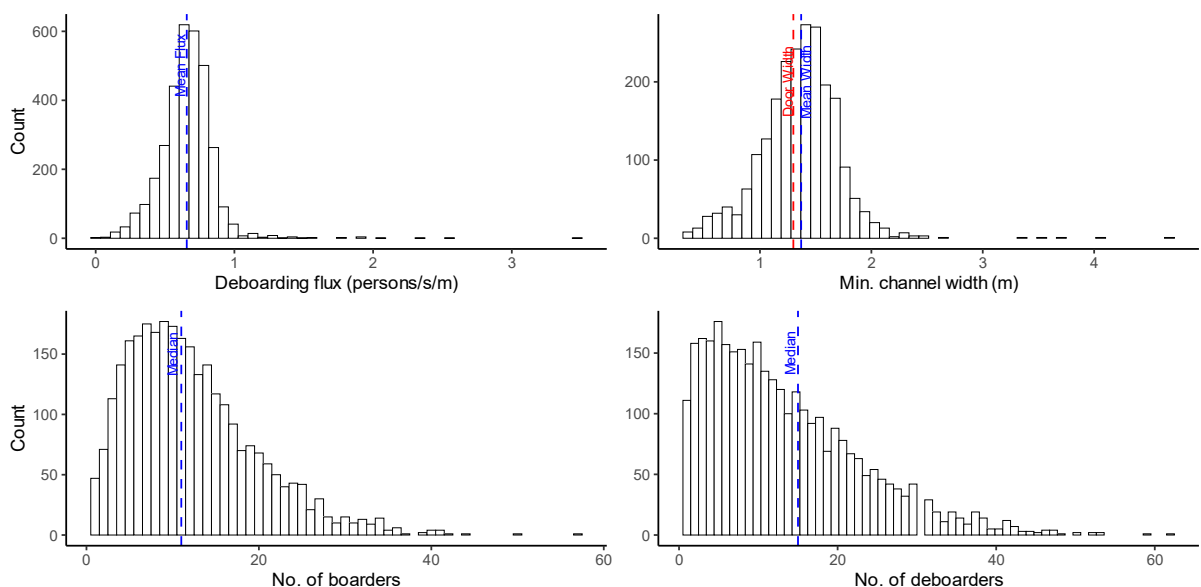

*Figure 3.* The figure shows histograms for deboarding flux (*persons/s/m*), channel width (*m*), number of boarders, and number of deboarders. The blue vertical lines in the top panels indicate the mean deboarding flux and mean channel width respectively. The red vertical line indicates the actual door width (*1.3 m*). The blue vertical lines in the bottom panels indicate the median number of boarders and deboarders.

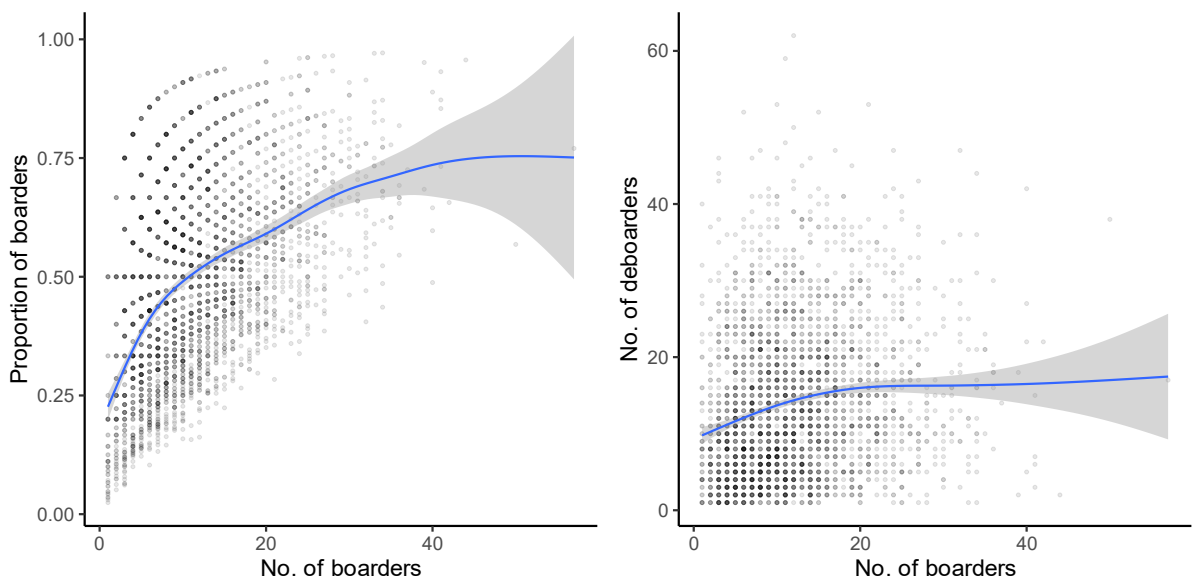

*Figure 4.* The figure on the left shows a positive correlation between the number of boarders (x-axis) and the proportion of boarders (y-axis). The figure on the right shows a small positive correlation between the number of boarders (x-axis) and the number of deboarders.

**Number of Boarders/Deboarders and Deboarding Flux**

First, we wanted to test whether an increase in the number of boarders and increase in proportion of boarders reduces the deboarding flux as shown by previous research (e.g., Seriani et al., 2016). To analyze this, we used a multilevel model with deboarding flux as the dependent variable, and number of boarders, number of deboarders, and proportion of boarders as fixed factor independent variables. We included the number of deboarders in the analysis to account for its effect on the deboarding flux. We included standardized timestamps as random intercepts to account for the variation across timeframes within a boarding event. The residuals of the model were normally distributed. In contrast to earlier research, the analysis showed that an increase in number of boarders increased the flux of deboarders ($B = 0.0063$, $SE = 0.00003$, $95\%\ CI[0.0062, 0.0064]$). This was a small effect with an increase of 0.06 *persons/s/m* for every 10 added boarders. The contrast with earlier research could be explained through the choice of variables such as total duration/flux of boarding and deboarding for previous analysis and not deboarding flux.

An increase in the number of deboarders also increased the deboarding flux ($B$ = 0.0004, $SE$ = 0.00003, *95% CI*[0.0003, 0.0005]). There was a very small increase of 0.004 *persons/s/m* for every 10 added deboarders. Figure 5 shows that there is only a small initial increase in flux with increasing deboarders after which the association becomes flat. In line with earlier research, we found that an increase in the proportion of boarders compared to deboarders also reduced the flux of deboarders ($B$ = -0.399, $SE$ = 0.002, *95% CI*[-0.402, -0.395]). For every increase in proportion of 0.1 (an increase of one-tenth in the proportion of boarders) the deboarding flux reduces by 0.04 *persons/s/m.* This result was consistent across groups with different number of boarders, thus smaller or larger boarding groups did not change these results. Results are visualized in Figures 5 and 6.

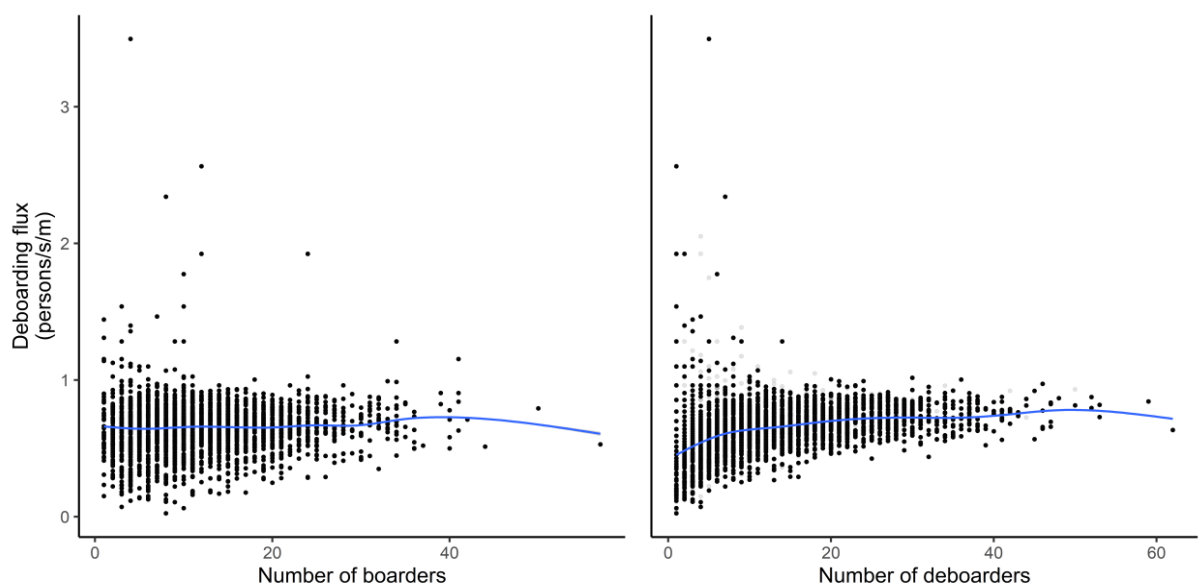

*Figure 5.* The plots show the relationship between the number of boarders and the deboarding flux (left) and the relationship between the number of deboarders and deboarding efficiency. The x-axis denotes the number of boarders (left), number of deboarders (right) and the y-axis denotes the deboarding flux (in *persons/s/m*). In both plots, the black dots denote the raw data points and the blue line is the fitted regression line.

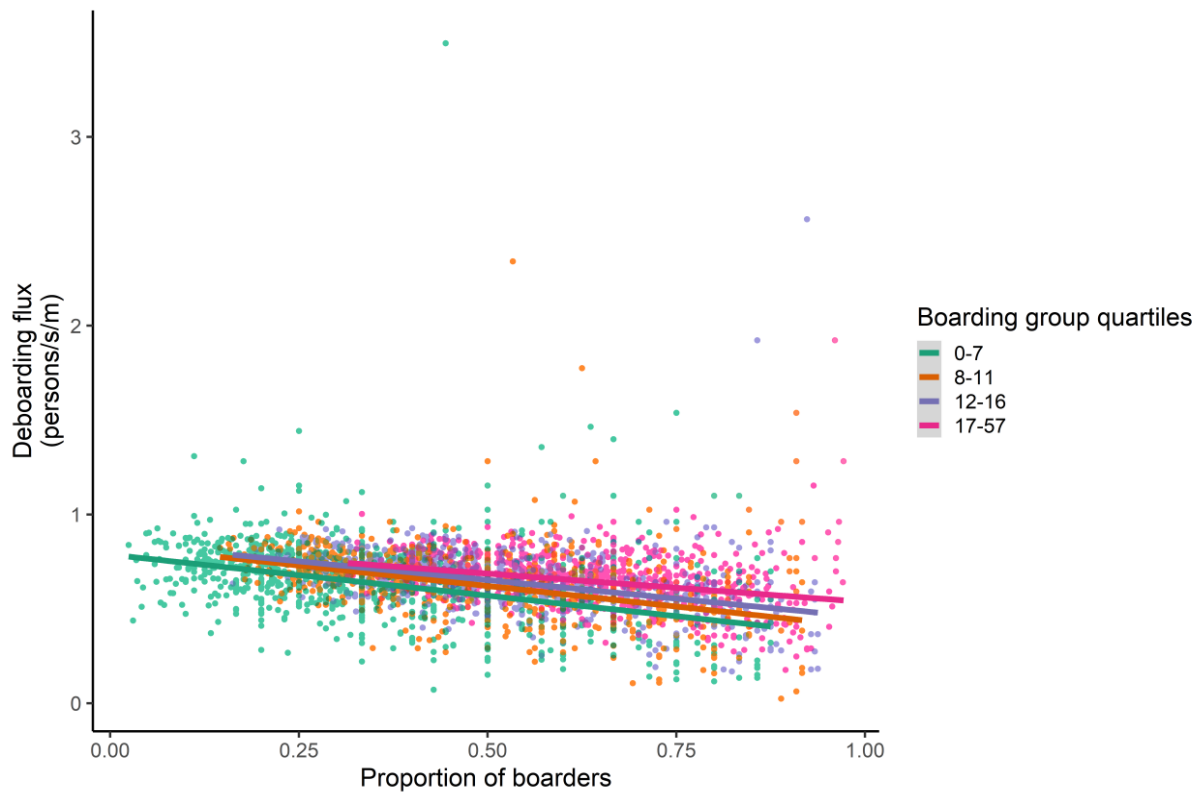

*Figure 6.* The plot shows the relationship between the proportion of boarders and deboarding efficiency. The x-axis denotes the proportion of boarders at a door and the y-axis denotes deboarding flux (in *persons/s/m*). The dots indicate the raw data and the lines are fitted regression lines. The colors indicate different quartiles of boarding groups with different number of boarders.

**Number of Boarders/Deboarders and Channel Width**

Next, we tested whether an increase in the number of boarders and proportion of boarders reduced the width of the channel formed by the group of waiting boarders. To analyze this, we used a multilevel model with channel width *(in meters)* as the dependent variable, and number of boarders, number of deboarders, and proportion of boarders as the fixed factor independent variables. We included the number of deboarders in the analysis to account for its effect on the channel width. We included standardized timestamps (*in s*) as random intercepts to account for the variation across timeframes within a boarding event. The residuals of the model were normally distributed. As expected, we found that an increase in the number of boarders reduced the width of the channel (*B* = -0.0102, *SE* = 0.0001, *95% CI*[-0.0105, -0.0099]). For every increase of 10 boarders, the channel width reduced by 10 *cm*. In Figure 7, we also observe that the variation in the channel width lowers as the number of boarders increase. We found that an increase in the number of deboarders increased the channel width ($B$ = 0.0047, $SE$ = 0.0001, *95% CI*[0.0045, 0.0049]). For every increase of 10 deboarders, the channel width increased by 4 cm. We also found that an increase in the proportion of boarders to deboarders reduced the channel width ($B$ = -0.22, $SE$ = 0.0081, *95% CI*[-0.23, -0.20]). This means that for every 0.1 or one-tenth increase in proportion of boarders, the channel width reduces by 2 cm. In Figure 8 we observe that these results remain fairly consistent for varying number of boarders. These results indicate that both boarders and deboarders influence the channel width in opposite directions thus implying that both groups exert a force on each other dynamically forming the deboarding channel.

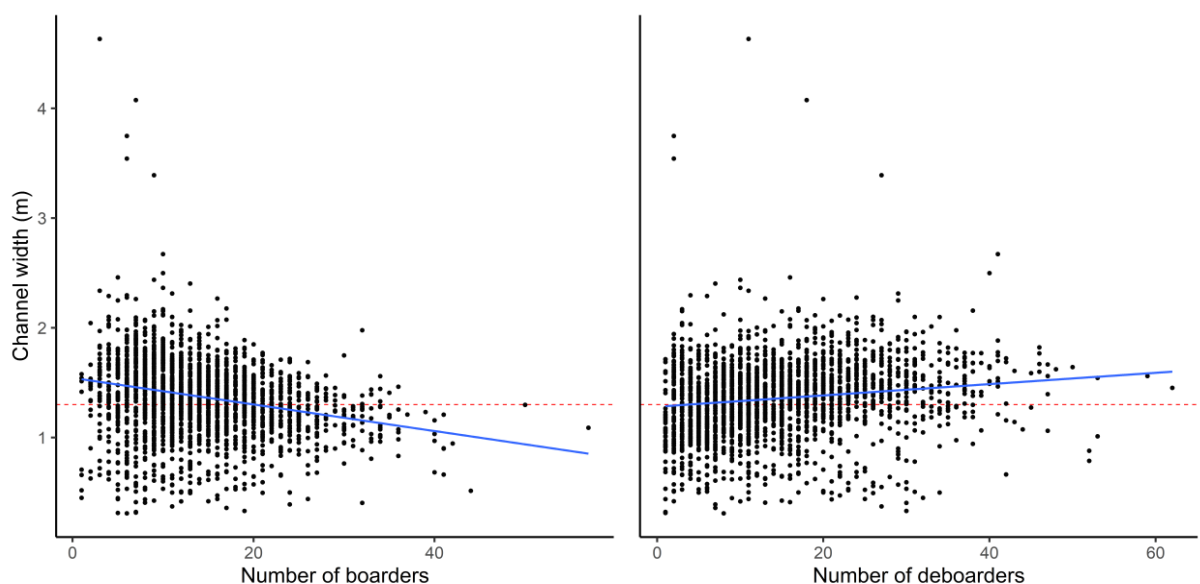

*Figure 7.* The plots show the relationship between the number of boarders (left) and the width of the deboarding channel and the number of deboarders (right) and the width of the channel. The x-axis denotes the number of boarders (left) and number of deboarders (right) and the y-axis denotes the channel width (in *m*). The black dots indicate the raw data points and the blue line is the fitted regression line. The red line indicates the train door width of 1.3 *m*.

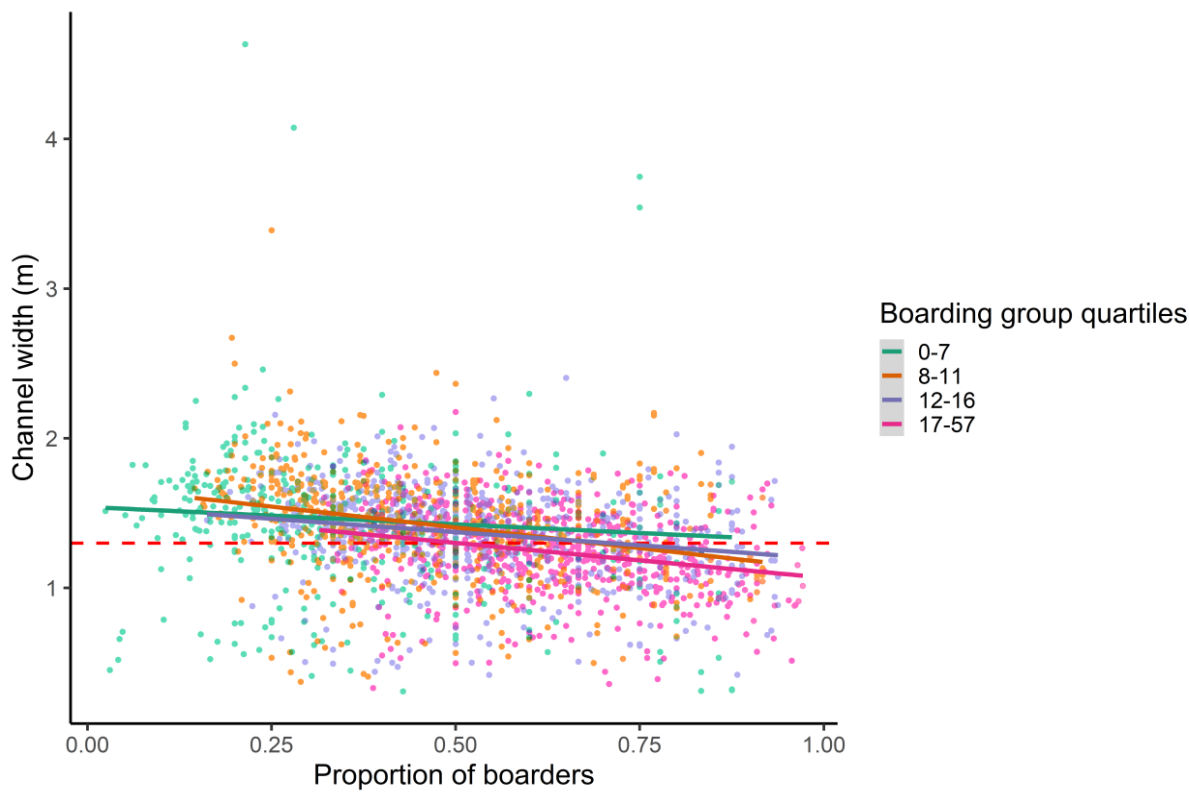

*Figure 8.* The plot shows the relationship between the proportion of boarders and the width of the deboarding channel. The x-axis denotes the proportion of boarders at a door and the y-axis denotes channel width (in *m*). The dots indicate the raw data and the lines are fitted regression lines. The colors indicate different quartiles of boarding groups with different number of boarders. The red line indicates the train door width of 1.3 *m*.

**Channel Width and Deboarding Flux**

Finally, we tested the effect of the channel width on the flux of the deboarders (see Figure 9). We expected that lower channel width would lower the deboarding flux. To analyze this, we used a multilevel model with deboarding flux *(in persons/s/m)* as the dependent variable, and channel width (*in meters*) as the fixed factor independent variable. We included standardized timestamps as random intercepts to account for the variation across timeframes within a boarding event. The residuals of the model were normally distributed. We found that when the channel width is larger, the flux of deboarding increases ($B = 0.092$, $SE = 0.0006$, $95\%\ CI[0.091, 0.093]$). This means that for every increase of 10 cm in channel width, the deboarding flux increases by ~ 0.01 *persons/s/m*

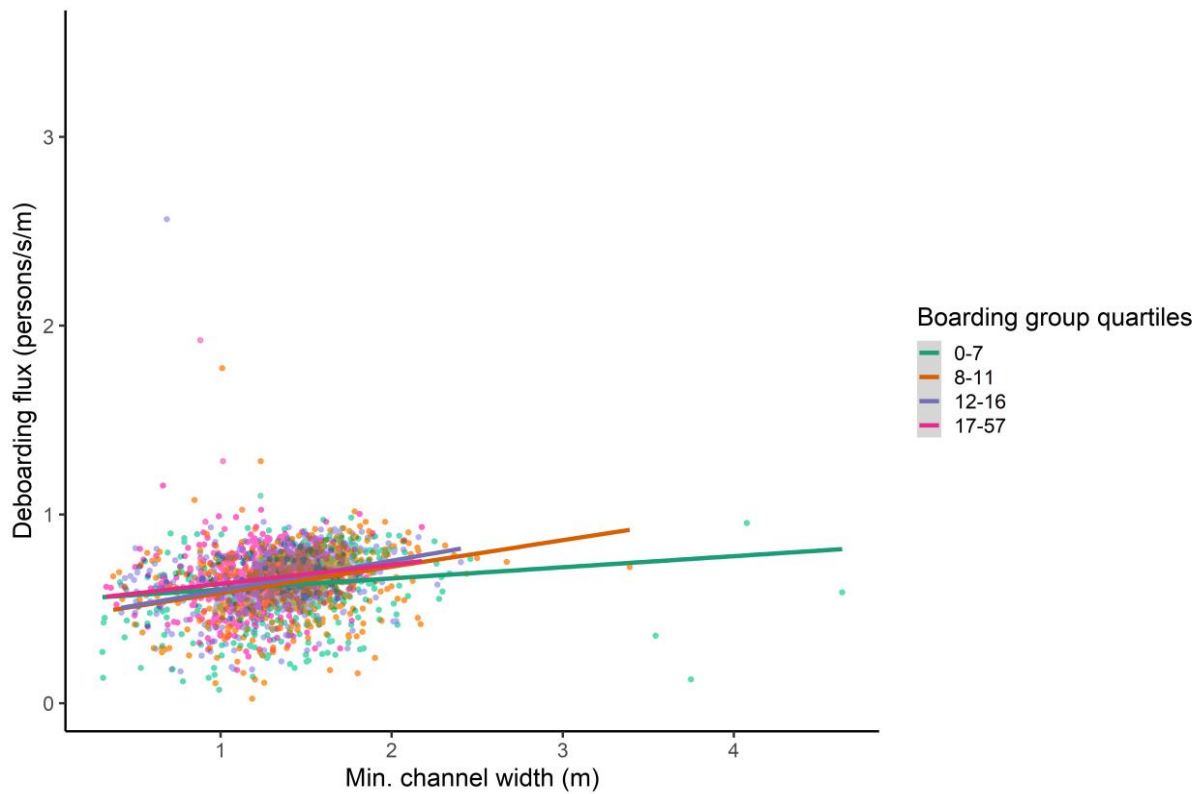

*Figure 9.* The plot shows the relationship between the width of the deboarding channel and the deboarding efficiency. The x-axis denotes the channel width (in *m*) at a door and the y-axis denotes the deboarding flux (in *persons/s/m*). The dots indicate the raw data and the lines are fitted regression lines. The colors indicate different quartiles of boarding groups with different number of boarders.

## General Discussion

The main question that this research answers is whether the width of the deboarding channel – a macroscopic structure formed as a result of the interaction between train boarders and deboarders – is influenced by the size of these groups. It also assesses whether this channel width in turn affects the efficiency of the deboarding process.

With real-life data from Utrecht Centraal station in the Netherlands, we found that the higher the number of waiting boarders and the higher the proportion of waiting boarders, the lower the width of the channel formed for deboarders to exit the train. Additionally, the higher the number of deboarders, the higher the width of the channel. We also found that the width of the deboarding channel was positively associated with the deboarding flux. Thus, when the width of the channel is lower, the flux of the deboarders is also lower, negatively affecting the deboarding efficiency. These results point to the dynamic effect that both boarders and deboarders have within this social system on a macroscopic structure such as the deboarding channel.

These results go beyond earlier research that shows that the number of boarders and proportion of boarders affects overall boarding/deboarding efficiency (de Ana Rodriguez et al., 2015; Olsson & Haugland, 2004; Seriani et al., 2016). They focus closely on deboarding efficiency and provide an explanation as to how efficiency is affected through the formation of macroscopic structures when the boarding and deboarding groups interact. These results imply that even though boarders are not actively trying to get into the train while the deboarding process is going on, the waiting boarders still have an influence on the efficiency of the deboarding process. As we argued earlier, these findings confirm that even though a social norm of waiting is present, it is not executed optimally in terms of deboarding efficiency. In other words, while most boarders adhere to the social norm of letting people deboard before boarding themselves, the space they leave for deboarders to exit the train is often sub-optimal.

One potential explanation for these results is that when there are higher numbers of boarders present at the door of the train, there might not be enough space on the platform to stand and so waiting boarders move closer to the door and take up space within the channel to create more space for the waiting boarders. However, the Voronoi area calculations show that the boarders on the outer edges of the waiting group of boarders have higher Voronoi areas indicating that they have more space available behind them (see Fig. 1). There is also space between the two doors of the same train carriage, but people do not tend to wait along the sides of the train. Instead, they prefer forming a semi-circular structure in front of the door forming a bulk queue (as was also found in other studies, see Kneidl et al., 2016; Wang et al., 2014).

Since space at the platform does not seem to be an issue, another more likely explanation is that when the number of waiting boarders is higher, boarders might come closer to the channel and make the channel smaller so that they can maximize their own chances of getting into the train early by preventing other boarders from moving in front of them. People might do this out of self-interest so that, for example, their chances of getting a seat are higher. But boarders could also do this to prevent other boarders from breaking the norm and getting on the train before the deboarders leave. Since this is a social norm, it is likely considered unfair or wrong to enter the train before deboarding is complete. Psychological research on cooperation has shown that punishment for those who do not cooperate is more effective when people interact with the same group repeatedly compared to a one-time interaction (Balliet et al., 2011). Because of its rather anonymous nature, the boarding process is effectively a one-time interaction—a one shot game in game theory terminology—which makes it hard to punish boarders who might break the social norm. As a result, boarders may instead try to prevent this behavior and as a consequence narrow the width of the deboarding channel.

**Implications and Future Research**

The results have implications for researchers who work on understanding the efficiency of the boarding process but also for practitioners who are involved in designing platforms and signage, and those who implement behavioral nudges or interventions to improve efficiency. These findings remove the focus from individual behavior change to identifying structures that can be changed when agents or groups of agents within social systems interact with each other. These results say more about the process of deboarding and how its efficiency is affected rather than testing only broader predictors such as the number of boarders and deboarders.

These findings provide insights on potential target areas on the platforms where the train doors are likely to be when the train stops at the platform. A clear indication of where boarders must wait when deboarders are exiting could be tested as a potential intervention to improve deboarding efficiency. Since an increasing number of boarders and higher proportion of boarders is negatively associated with corridor width, interventions can also focus on more equal distribution of boarders across the platform. The doors with the highest number of boarders would be the least efficient door in terms of boarding and deboarding, thus distribution of boarders would reduce the number of boarders present at the busiest doors, reducing the likelihood of a narrower deboarding corridor.

Future research can further build on these findings by empirically testing what individuals do—and why they do it—that results in narrowing of the channel width. Understanding the motivations and strategies of train boarders can tell us more about how individuals trying to maximize their efficiency might minimize the efficiency of the deboarders. Theories related to how people interact in cooperation situations such as interdependence theory (van Lange & Balliet, 2015) can be applied to crowd behavior and help in explaining the underlying motivations and strategies specific to the train boarding situation. This can be studied via focus groups or interviews with regular commuters. This can also be studied via agent-based models, where parameters at the micro level (at the level of the individual) can be varied to see how they affect the emergence of the channel. This method would then help us understand what parameters at the micro level affect the formation of macrolevel structures, especially if these micro-level interactions non-linearly influence the emergence of the deboarding channel.

## Conclusion

To summarize, we found that the size of the boarding and deboarding group influences the formation of the deboarding channel, a macroscopic structure formed as a result of the interaction between boarders and deboarders. Larger boarding groups lower the width of the channel, and lower channel widths in turn lower the efficiency with which deboarders exit the train, thus overall increasing the duration of deboarding and contributing to an increase in the dwell time of the train. These results imply that researchers and practitioners should target this macroscopic structure when designing behavior change interventions to improve the boarding and deboarding process.